\documentclass[aps,prb,showpacs,twocolumn,floats]{revtex4}
\usepackage{amssymb}
\usepackage{amsbsy}
\usepackage{amsmath}
\usepackage{epsfig}

\begin{document}

\title {Mott phases and quantum phase transitions of extended Bose-Hubbard models
in 2+1 dimensions}

\author{K. Sengupta}
\affiliation{TCMP Division, Saha Institute of Nuclear Physics, 1/AF
Bidhannagar, Kolkata-700064, India}

\date{\today}

\begin{abstract}

We review the recent theoretical developments towards understanding
the Mott phases and quantum phase transition of extended
Bose-Hubbard models on lattices in two spatial dimensions . We
focuss on the description of these systems using the dual vortex
picture and point out the crucial role played by the geometry of
underlying lattices in determining the nature of both the Mott
phases and the quantum phase transitions. We also briefly compare
the results of dual vortex theory with quantum Monte Carlo results.

\end{abstract}


\maketitle

\section {Introduction}

The superfluid to Mott insulators transitions of strongly correlated
lattice bosons systems, described by extended Hubbard models, in two
spatial dimensions have recently received a great deal of
theoretical interest. One of the reasons for this renewed attention
is the possibility of experimental realization of such models using
cold atoms trapped in optical lattices \cite{bloch1, kasevitch1}.
However, such transitions are also of interest from a purely
theoretical point of view, since they provide us with a test bed for
exploring the recently developed theoretical paradigm of
non-Landau-Ginzburg-Wilson (LGW) phase transitions \cite{senthil1}.
In the particular context of lattice bosons in two spatial
dimensions, a general framework for such non-LGW transitions has
been developed and applied to the case of square lattice
\cite{balents1}. The application of this framework to triangular
\cite{burkov1}, kagome \cite{sengupta1} lattices has also been
carried out. These theoretical works has also supplemented by
quantum Monte Carlo (QMC) studies which serves as a rigorous test
for predicted theoretical results. Such studies, for extended Bose
Hubbard models, has been carried out for square \cite{squaremonte},
triangular \cite{melko1,damle1}, kagome \cite{isakov1,damle2}, and
hexagonal lattices \cite{wessel1}. In this review, we present a
brief overview of the theoretical ideas developed in the
above-mentioned works focussing on role of the geometry of the
underlying lattices in determining the nature of both the Mott
phases and the quantum phase transitions. We also briefly compare
the results of dual vortex theory with QMC results.

The typical paradigm of non-LGW transitions that has been proposed
in the context of lattice bosons in Refs.\
\onlinecite{balents1,burkov1,sengupta1,burkov2} is the following. It
is proposed that for non-integer rational fillings $f=p/q$ of bosons
per unit cell of the underlying lattice ($p$ and $q$ are integers),
the theory of phase transition from the superfluid to the Mott
insulator state is described in terms of the vortices which are
non-local topological excitations of the superfluid phase, living on
the dual lattice.\cite{fisher1, tesanovic1} These vortices are not
the order parameters of either superfluid or Mott insulating phases
in the usual LGW sense. Thus the theory of the above mentioned phase
transitions are not described in terms of the order parameters on
either side of the transition which is in contrast with the usual
LGW paradigm of phase transitions. Also, as first explicitly
demonstrated in Refs.\ \onlinecite{balents1}, although these
vortices are excitations of a featureless superfluid phase, they
exhibit a quantum order which depends on the filling fraction $f$.
It is shown that the vortex fields describing the transition form
multiplets transforming under projective symmetry group (projective
representations of the space group of the underlying lattice and
that this property of the vortices naturally and necessarily
predicts broken translational symmetry of the Mott phase, where the
vortices condense. Since this translational symmetry breaking is
dependent on the symmetry group of the underlying lattice, geometry
of the lattice naturally plays a key role in determining the
competing ordered states of the Mott phase and in the theory of
quantum phase transition between the Mott and the superfluid phases.

In what follows, we shall develop the above-mentioned ideas
concentrating on square and Kagome lattices. The relevant results
for the triangular lattice can be found in Ref.\
\onlinecite{burkov1}. To this end, we consider the extended
Bose-Hubbard Hamiltonian
\begin{eqnarray}
H_{\rm boson}&=& -t\sum_{\left<ij\right> }\left( b_i^{\dagger} b_j +
{\rm h.c.} \right) +\frac{U}{2} \sum_i n_i\left(n_i-1\right) \nonumber\\
&& + V \sum_{\left<ij\right>} n_i n_j  -\mu \sum_i n_i.
\label{boson1}
\end{eqnarray}
Here $t$ is the boson hopping amplitude between nearest neighbor
sites, $U$ is the on-site interaction, $V$ denotes the strength of
the nearest neighbor interaction between the bosons and $\mu$ is the
chemical potential. We note at this point that a simple
Holstein-Primakoff transformations maps this boson model (Eq.\
\ref{boson1}), in the limit of hardcore bosons ($U \rightarrow
\infty$), to XXZ models with ferromagnetic $J_x$ and
antiferromagnetic $J_z$ interaction in a longitudinal magnetic field
$B_l$
\begin{eqnarray}
H_{\rm XXZ} &=& -J_x \sum_{\left<ij\right>} \left( S_i^x S_j^x +
S_i^y
S_j^y \right) + J_z \sum_{\left<ij\right>} S_i^z S_j^z \nonumber\\
&& -B_l \sum_i S^z_i \label{spinmodel1}
\end{eqnarray}
where $J_x >0 $ and $J_z> 0$ are the strengths of transverse and
longitudinal nearest neighbor interactions and $B_l$ is a
longitudinal magnetic field. The parameters $J_x$, $J_z$ and $B_l$
of $H_{\rm XXZ}$ can be mapped to those of $H_{\rm boson}$: $J_{x}=
2t$, $J_z= V$, and $B_l = zV\left(f-\frac{1}{2}\right)$, where $z$
denotes the coordination number of the underlying lattice and $f$ is
the average boson filling.

\section {Competing Mott states and quantum phase transition}
\label{Mott}

The basic premise underlying the theories developed in Refs.\
\onlinecite{balents1,burkov1,sengupta1,burkov2} is that the quantum
phase transition of the extended Bose-Hubbard model occurs due to
destabilization of the superfluid phases by proliferation of
vortices which are topological excitations of the superfluid phase.
To analyze such a transition, one therefore needs to obtain an
effective action for the vortex excitation. Such an action can be
obtained by performing a duality analysis of the Bose-Hubbard model
\cite{dasgupta1,balents1}. The final form of the dual action $S_d$
can be written as \cite{balents1}
\begin{eqnarray}
S_d &=& \frac{1}{2e^2} \sum_b \left(\epsilon_{\mu \nu \lambda}
\Delta_{\nu} A_{b\lambda} - f\delta_{\mu \tau} \right)^2 \nonumber\\
&& - y_v \sum_b \left(\psi_{b +\mu} e^{2\pi i A_{b \mu}} \psi_b +
{\rm h.c.} \right) \nonumber\\
&&  + \sum_b \left( r \left|\psi_b \right|^2 + u \left|\psi_b
\right|^4 \right) \Bigg]
\end{eqnarray}
where $\psi_b$ are the vortex field living on the site $b$ of the
dual lattice, $A_{b\mu}$ is the U(1) dual gauge field so that
$\epsilon_{\tau \nu \lambda} \Delta_{\nu} A_{b \lambda}= n_i$ where
$n_i$ is the physical boson density at site $i$, $\sum_p$ denotes
sum over elementary plaquette of the dual lattice$, \Delta_{\mu}$
denotes lattice derivative along $\mu=x,y,\tau$, and $f$ is the
average boson density. Note that the vortex action $S_d$ is not
self-dual to the boson action obtained from $H_{\rm boson}$.
Therefore we can not, in general, obtain a mapping between the
parameters of the two actions, except for identifying
$\epsilon_{\tau \nu \lambda} \Delta_{\nu} A_{b\lambda}$ as the
physical boson density \cite{balents1}. We shall therefore classify
the phases of this action based on symmetry consideration and within
the saddle point approximation as done in Refs.\
\onlinecite{balents1,burkov1,sengupta1,burkov2}.

The transition from a superfluid ($\left<\psi_b \right> =0$) to a
Mott insulating phase in $S_d$ can be obtained by tuning the
parameter $r$. For $r>0$, we are in the superfluid phase. Note that
the saddle point of the gauge fields $A_{b\mu}$ in action
corresponds to $\epsilon_{\tau \nu \lambda} \Delta_{\nu} {\bar
A}_{b\lambda}= f$, so that the magnetic field seen by the vortices
is pinned to the average boson filling $f$. Now as we approach the
phase transition point $r=0$, the fluctuations about this saddle
point ($\left<\psi_b \right> =0$, ${\bar A}_{by} = f x$) increase
and ultimately destabilize the superfluid phase in favor of a Mott
phase with $\left<\psi_b \right> \ne 0$. Clearly, in the above
scenario, the most important fluctuations of the vortex field
$\psi_b$ are the ones which has the lowest energy. This prompts us
to detect the minima of the vortex spectrum by analyzing the kinetic
energy term of the vortices $H_{\rm kinetic} = -y_v \sum_{b,\alpha }
\left(\psi_{b+\alpha} ^{\dagger} e^{2\pi i {\bar A}_{b}} \psi_{b} +
{\rm h.c} \right)$, where the sum over $\alpha$ is carried out over
the sites of the dual lattice which are nearest neighbors to $b$.
The analysis of $H_{\rm kinetic}$ therefore amounts to solving the
Hofstadter problem on the dual lattice which has been carried out in
Refs.\ \onlinecite{balents1}, \onlinecite{burkov1}, and
\onlinecite{vidal1,ye1,sengupta1} for square, triangular and Kagome
lattices respectively.

\subsection{Square lattice}
\label{square}

The dual lattice in this case is also a square lattice with its
sites being at the center of the plaquettes of the direct lattice.
Thus here one needs to solve the Hofstadter problem on a square
lattice with $f=p/q$ flux quanta per plaquette of the dual square
lattice. It has been shown in Ref.\ \onlinecite{balents1} that for
$f=p/q$, there are $q$ minima within the magnetic Brillouin zone at
the wave-vectors ${\bf Q}= (k_x a,k_y a) =(0,2\pi l p/q)$, where $l$
runs from $0$ to $q-1$, $k_{x,y}$ are the wavevectors of the
magnetic Brillouin zone and $a$ is the lattice spacing. The
eigenfunctions $\psi(k_x,k_y)$ corresponding to these minima can be
easily constructed \cite{balents1,senthil1}.

Once the positions of these minima are located, the next task is to
identify the low lying excitations around these minima which are
going to play a leading role in destabilizing the superfluid phase.
These excitations, represented by the bosonic fields $\varphi_l$,
has definite transformation properties under the symmetry operations
of the underlying dual lattice \cite{balents1}. For the case of
square lattice, the relevant symmetry operations are translations
along $x$ and $y$ directions by one lattice spacing ($T_x$ and
$T_y$), reflections about $x$ and $y$ axes ($I_x$ and $I_y$) and
rotation by $\pi/2$ ($R_{\pi/2}$). It has been shown in Ref.\
\onlinecite{balents1} that the transformation properties of
$\varphi_l$ under these transformations are the following: $T_x:
\,\varphi_{l} \rightarrow \varphi_{l-1}$, $T_y: \, \varphi_{l}
\rightarrow \varphi_l \omega^{l}$, $R_{\pi/2}: \, \varphi_l
\rightarrow \frac{1}{\sqrt{q}} \sum_{l'=0}^{q-1} \omega^{ l l'}
\varphi_{l'}$, $I_x: \, \varphi_l \rightarrow \varphi_{l}^{\ast}$,
and $I_y: \, \varphi_l \rightarrow \varphi_{-l}^{\ast}$, where
$\omega = \exp(-2\pi i f)$.

Finally, we need to construct an effective LGW action in terms of
the low-energy fields $\varphi_l$ which is invariant under all the
symmetry operations. For the sake of definiteness, we shall focuss
on the case $q=2$ and $q=3$ from now on. First, let us consider the
case $q=2$ for which, the effective action reads
\cite{balents1,senthil2},
\begin{eqnarray}
S_v &=& \int d^2 r dt (L_2 + L_4 +L_8)  \nonumber\\
L_2 &=& \left( \sum_{\ell = 0}^{1} \left[ |(\partial_\mu - i A_{\mu}
) \varphi_l |^2 + s |\varphi_l |^2 \right] + \frac{1}{2e^2} \left(
\epsilon_{\mu\nu\lambda}
\partial_\nu A_\lambda \right)^2 \right). \nonumber\\
L_4 &=& \frac{\gamma_{00}}{4} \left( |\varrho_0|^2 + |\varrho_1|^2
\right)^2 - \frac{\gamma_{01}}{4} \left( |\varrho_0|^2 - |
\varrho_1|^2 \right)^2 \nonumber\\
L_8 &=&  \lambda (\varrho_0 \varrho_1^\ast )^4 + {\rm h.c}.
\label{q2lang}
\end{eqnarray}
where $ \varrho_{0(1)}= \left(\varphi_0 \pm \varphi_1\right)/{\sqrt
2}$. Note that we have included a $8^{\rm th}$ order term $L_8$ in
the effective action since this is the lowest order term, invariant
under all the symmetry operations, which determines the relative
phase of the vortex fields $\varrho_0$ and $\varrho_1$.

The Mott phases, which are stabilized for $s < 0$, can be obtained
for $q=2$ by a straightforward analysis of $L_4$ and $L_8$. To
concisely present the results for these phases, it is convenient to
define a generalized density $\delta \rho ({\bf r}) =
\sum_{m,n=-1,1} \rho_{mn} e^{i(m r_x + n r_y)}$, where $\rho_{mn}=
S\left(|{\bf Q}_{mn}|\right ) \omega^{mn/2} \sum_{l = 0}^{1}
\varphi^{\ast}_l \varphi_{l+n} \omega^{l m}$ denotes the most
general gauge-invariant bilinear combinations of the vortex fields
$\varphi_l$ with appropriate transformation properties of square
lattice space group \cite{balents1}. As shown in Ref.\
\onlinecite{balents1}, the values of $\delta \rho ({\bf r})$ on
sites of the dual lattice (integer $r_x$ and $r_y$) can be
considered a measure of the ring-exchange amplitude of bosons around
the plaquette whereas those with ${\bf r}$ half-odd-integer
co-ordinates represent sites of the direct lattice, and the values
of $\delta \rho ({\bf r})$ on such sites measure the boson density
on these sites. Finally, ${\bf r}$ values with $r_x$ integer and
$r_y$ half-odd-integer correspond to vertical links of the square
lattice (and vice versa for horizontal links), and the values of
$\delta \rho ({\bf r})$ on the links is a measure of the mean boson
kinetic energy; if the bosons represent a spin system, this is a
measure of the spin exchange energy. We also note two fundamental
points from the definition of $\rho_{mn}$. First, as long as
$\langle\varphi_l \rangle \ne 0$ for at least one non-zero ${\bf
Q}_{mn}$, the Mott state is characterized by a non-trivial density
wave order. Second, the relative phase of the boson fields
$\varrho_l$ which plays a crucial role in determining the nature of
the density wave order is fixed by the $8^{\rm th}$ order term
$L_8$.
\begin{figure}
\rotatebox{0}{
\includegraphics[width=0.9\linewidth]{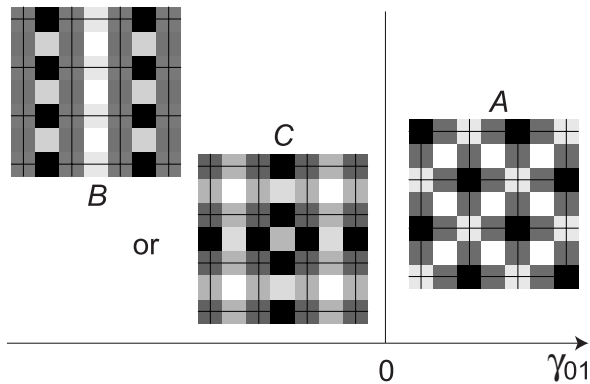}}
\caption{Plot of $\delta \rho ({\bf r})$ on the sites, links and
plaquettes of the direct lattice: the lines represent links of the
direct lattice. As discussed in the text, these values represent the
boson density, kinetic energy, and ring-exchange amplitudes
respectively. The choice between the states $B$ and $C$ is made by a
eighth order term in the action.} \label{fig1}
\end{figure}

The Mott phases for $q=2$ is shown in Fig.\ \ref{fig1}. We note that
there are three possible phases. For $\gamma_{01}>0$, only one of
the two vortex fields condenses and the resulting Mott state (state
A in Fig.\ \ref{fig1}) has a charge density wave order. For
$\gamma_{01} <0$, both the vortex fields condense and this leads to
two possible Mott states (states B and C in Fig.\ \ref{fig1}). For
these states, all sites of the direct lattice are equivalent. These
therefore represent valence bond solid (VBS) states with columnar or
plaquette VBS order parameters.

Next, we address the issue of quantum phase transition between the
superfluid and the Mott state for $q=2$. Such a transition, within
the premise of dual vortex theory discussed here, is described by
$S_v$ and not, as would be expected by LGW paradigm, by an action
written in terms of order parameters fields at either side of the
transition. The phase transition can either be first or second
order. If it turns out to be second order and if $v <0$, the
relative phase of the vortex fields, which is pinned in the Mott
phase by the dangerously irrelevant (in the renormalization group
sense) $8^{\rm th}$ order term $L_8$, becomes a gapless mode at the
transition \cite{balents1}. Thus such a transition has an emergent
gapless mode at the critical point. It provides an example of a
deconfined quantum critical point \cite{senthil1} and can be shown
to accompanied by boson fractionalization \cite{balents1}.

\begin{figure}
\rotatebox{0}{
\includegraphics[width=0.9\linewidth]{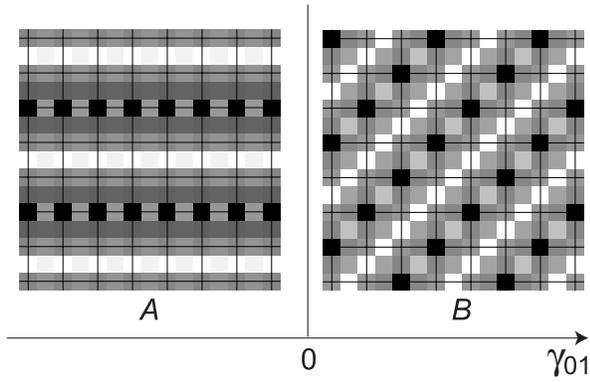}}
\caption{Plot of $\delta \rho ({\bf r})$ for $q=3$.} \label{fig2}
\end{figure}

The Mott phases for $q=3$ can also be similarly obtained
\cite{balents1}. Here the quartic term in the effective Lagrangian
which determines the Mott states are given by
\begin{eqnarray}
\mathcal{L}_4 &=& \frac{\gamma_{00}}{4} \left( |\varphi_0|^2 +
|\varphi_1|^2  + |\varphi_2|^2 \right)^2 \nonumber \\ &+&
\frac{\gamma_{01}}{2} \left( \varphi_0^\ast \varphi_1^\ast
\varphi_2^2 + \varphi_1^\ast \varphi_2^\ast \varphi_0^2 +
\varphi_2^\ast \varphi_0^\ast \varphi_ - \varphi_0^\ast
\varphi_1^2 + \mbox{c.c.} \right. \nonumber \\
&-& \left. 2 |\varphi_0|^2 |\varphi_1|^2 - 2 |\varphi_1|^2
|\varphi_2|^2- 2 |\varphi_2|^2 |\varphi_0|^2\right). \label{m6}
\end{eqnarray}
Note that here the ordering of the Mott states are completely
determined by ${\mathcal L}_4$ and retaining higher order terms are
not essential. It turns out that there are two sixfold degenerate
VBS states with columnar or diagonal VBS orders as shown in Fig.\
\ref{fig2}. The superfluid-Mott insulator quantum phase transition
is again described by a dual vortex action, but does not lead to
deconfined quantum criticality as pointed out in Ref.\
\onlinecite{balents1}.

QMC studies on Bose-Hubbard models on square lattice with nearest
neighbor interaction and near half-filling has been carried out in
Ref.\ \onlinecite{squaremonte}. These studies indicate a
checkerboard Mott state and a strong first order transition between
the superfluid and checkerboard Mott state. Analogous QMC studies on
triangular lattice has also been carried out in Refs.\
\onlinecite{melko1,damle1}.

\subsection{Kagome lattice}

The dual lattice corresponding to Kagome is the well-known dice
lattice which has three inequivalent sites denoted commonly by $A$,
$B$ and $C$ \cite{vidal1}. To obtain the minima of the vortex
spectrum, we therefore need to solve the Hofstadter problem on the
dice lattice. Here we shall concentrate only for $f=1/2$ and $2/3$
(or equivalently $f=1/3$).

\begin{figure}
\rotatebox{0}{
\includegraphics[width=0.5\linewidth]{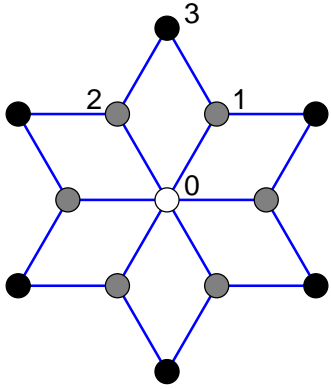}}
\caption{Aharanov-Bohm cages for vortices in a dice lattice. See
text for details.} \label{fig3}
\end{figure}

For $f=1/2$, it has been shown \cite{vidal1,ye1} that the vortex
spectrum on dice lattice do not have well-defined minima and
collapse to three degenerate bands. Physically, the collapse of the
vortex spectrum can be tied to localization of the vortex within the
so-called Aharanov-Bohm cages as explicitly demonstrated in Ref.\
\onlinecite{vidal1}. An example of such a cage is shown in Fig.\
\ref{fig3}. A vortex whose initial wavepacket is localized at the
central (white) $A$ site can never propagate beyond the black sites
which form the border of the cage. This can be understood in terms
of destructive Aharanov-Bohm interference: The vortex has two paths
$0\rightarrow1\rightarrow 3$ and $0\rightarrow 2\rightarrow 3$ to
reach the rim site $3$ from the starting site $0$. The amplitudes
from these paths destructively interfere for $f=1/2$ to cancel each
other. Thus the vortex remain within the cage. Such dynamic
localization of the vortex wavepackets, termed as Aharanov-Bohm
caging, makes it energetically unfavorable to condense vortices.
Thus superfluidity persists for arbitrarily small values of $t/V$.
In the language of spins, this also explains the absence of $S_z$
ordering for XXZ model in a Kagome lattice for $B_l=0$ and $J_z \gg
J_x$. Such a persistence of superfluidity has also been confirmed by
QMC studies \cite{isakov1,damle2}.

In contrast to $f=1/2$, the vortex spectrum has two well-defined
minima for $f=2/3$ within the magnetic Brillouin zone located at $
(k_x a, \kappa) = (0, \pi/3)\, {\rm and} \,(2\pi/3, 2 \pi/3)$, where
$\kappa = \sqrt{3}k_y a/2$. Thus the low energy properties of the
vortex system can be characterized in terms of the two low-energy
bosonic fields $\varphi_1$ and $\varphi_2$ similar to the case of
square lattice at $f=1/2$. However, as shown in Ref.\
\onlinecite{sengupta1}, in contrast to the case of the square
lattice, the space group of symmetry transformations of the dice
lattice involves translations by lattice vectors $u= ()$ and $v=()$
($T_u$ and $T_v$), rotation by $\pi/3$ ($R_{\pi/3}$), and
reflections around $x$ and $y$ ($I_x$ and $I_y$). The transformation
properties of the vortex fields under these symmetry operations are
the following: $T_{u}:\, \varphi_{1} \rightarrow \varphi_{1}
\exp(-i\pi/3),\, \varphi_{2} \rightarrow \varphi_{2} \exp(i\pi/3)$,
$T_{v}: \, \varphi_{1} \rightarrow \varphi_{1} \exp(i\pi/3),\,
\varphi_{2} \rightarrow \varphi_{2} \exp(-i\pi/3)$, $I_x:\,
\varphi_{1(2)} \rightarrow \varphi^{\ast}_{1(2)}$,
$I_y:\,\varphi_{1(2)} \rightarrow \varphi^{\ast}_{2(1)}$, and
$R_{\pi/3}: \varphi_{1(2)}\rightarrow \varphi_{2(1)}$
\cite{sengupta1}.

The simplest Landau-Ginzburg theory for the vortex fields which
respects all the symmetries is \cite{sengupta1}
\begin{eqnarray}
L_v &=& L_v^{(2)}+ L_v^{(4)}+ L_v^{(6)} \\
L_v^{(2)} &=& \sum_{\alpha=1,2} \left[ \left|\left(\partial_{\mu} -
e A_{\mu}\right) \phi_{\alpha} \right|^2 +
r \left|\varphi_{\alpha}\right|^2  \right]\\
L_v^{(4)} &=& u\left(\left|\varphi_1\right|^4 +
\left|\varphi_2\right|^4 \right) + v \left|\varphi_1\right|^2
\left|\varphi_2\right|^2 \\
L_v^{(6)} &=& w \left[ \left(\varphi_1^{\ast} \varphi_2 \right)^3 +
{\rm h.c.} \right]
\end{eqnarray}
Here we find that $L_v^{(6)}$ turns out to be the lowest-order term
which breaks the $U(1)$ symmetry associated with the relative phase
of the vortex fields.

\begin{figure}
\rotatebox{0}{
\includegraphics[width=0.9\linewidth]{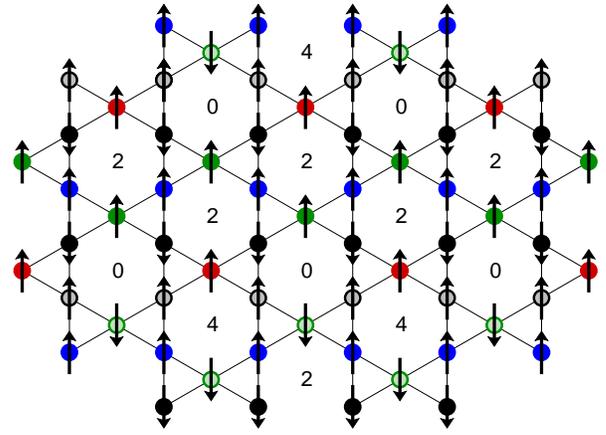}}
\caption{Density wave state with $9$ by $9$ order (full pattern not
shown) for $f=2/3$. Up/down arrows denotes singly occupied/empty
sites. See text for details.} \label{fig4}
\end{figure}

A simple power counting shows that $L_v^{(6)}$ is marginal at tree
level. Unfortunately, the relevance/irrelevance of such a term
beyond the tree level, is not easily determined analytically
\cite{sengupta1}. If it so turns out that $L_v^{(6)}$ is irrelevant,
the situation here will be identical to that of bosons on square
lattice at $f=1/2$. The relative phase of the vortices would emerge
as a gapless low energy mode at the critical point. The quantum
critical point would be deconfined and shall be accompanied by boson
fractionalization \cite{balents1}. On the other hand, if $L_v^{(6)}$
is relevant, the relative phase degree of the bosons will always
remain gapped and there will be no deconfinement at the quantum
critical point. Here there are two possibilities depending on the
sign of $u$ and $v$. If $u,v<0$ and $w>0$, the transition may become
weakly first order whereas for $u>0$, it remains second order (but
without any deconfinement). QMC studies seem to indicate a weak
first order transition in this case \cite{isakov1}. However, it is
possible that a second order quantum phase transition can still be
possible at special points on the phase diagram and this issue
remains to be settled \cite{isakov1,damle2,damle3}.

The vortices condense for $r<0$ signifying the onset of Mott phases.
It turns out that the Mott phase for $v>0$ do not lead to any
density wave order \cite{sengupta1}. For $v <0$, there are two
possible Mott phases as demonstrated in Ref.\
\onlinecite{sengupta1}. One of these phases with $9$ by $9$ ordering
pattern is shown in Fig.\ \ref{fig4}. Here the bosons are localized
in red, blue, green (closed circle) and black (open circle) sites
whereas the green (open circle) and black (closed circle) sites are
vacant leading to a filling of $f=2/3$. The net occupation of the
hexagons takes values $0$, $4$ and $2$ as shown in Fig.\ \ref{fig4}.
Note that interchanging the occupations of all the black and green
sites of this state (while leaving the red and the blue sites
filled) has the effect of $2\leftrightarrow 4$ for the boson
occupation of the hexagons labeled $2$ and $4$ in Fig.\ \ref{fig4}
while leaving those for hexagons labeled $0$ unchanged. The
resultant state is degenerate (within mean-field theory) to the
state shown in Fig.\ \ref{fig4}. This allows us to conjecture that
the effect of quantum fluctuations on the state is shown in Fig.\
\ref{fig4} might lead to stabilization of a partial resonating state
with a $3$ by $3$ $R-3-3$ ordering pattern \cite{sengupta1,burkov2}.
Such a state is indeed found to be the true ground state in exact
diagonalization \cite{cabra1} and QMC studies \cite{isakov1}. Note
that the mean-field analysis of vortex theory which neglects quantum
fluctuations can not directly give this partially resonating state.

\section{Discussion}
\label{conclusion}

In conclusion, we have briefly reviewed theoretical developments on
duality analysis of Bose-Hubbard and equivalently XXZ models on two
dimensional lattices. We have focussed mainly at predictions such an
analysis for square and Kagome lattices at half and one-third
filling. We have compared the predictions of such dual vortex
theories with QMC results wherever possible. Finally, we would like
to point out a couple of issues that still remain unsettled in this
regard. It is yet unknown, whether it is possible to have a second
order quantum phase transition in these models as predicted by the
vortex theory. Direct QMC studies
\cite{squaremonte,damle1,melko1,isakov1,damle2,damle3} as well as
several other studies \cite{kuklov1} seem to find at best a weak
first order transition. However, it has been recently conjectured
that at least for the Kagome lattice, a second order transition
might occur at special particle-hole symmetric points in the phase
diagrams \cite{damle3}. Since QMC studies, at least for large enough
system sizes, can only reach close to this point, this issue is not
settled yet. Second, it remains to be seen whether such duality
analysis for boson theories with $U(1)$ symmetry can be extended to
either models with Fermions and/or bosonic models with higher
symmetry group.

The author thanks L. Balents, L. Bartosch, A. Burkov, S. Isakov,
Y.B. Kim, R. Melko, S. Sachdev, and S. Wessel for collaborations on
related earlier works, and  K. Damle, T. Senthil, and A. Vishwanath
for helpful discussions.


\vspace{-0.6 cm}

\begin{thebibliography}{99}

\bibitem{bloch1} M. Greiner, O. Mandel, T.Esslinger, T.W. Ha¨nsch,
and I. Bloch, Nature (London) {\bf 415}, 39 (2002).

\bibitem{kasevitch1} C. Orzel, A.K. Tuchman, M.L. Fenselau, M. Yasuda,
and M.A. Kasevich, Science {\bf 291}, 2386 (2001).

\bibitem{senthil1} T. Senthil {\it et\, al.}, Science {\bf 303}, 1490
(2004); T. Senthil {\it et\, al.}, Phys. Rev. B, {\bf 70}, 144407
(2004).

\bibitem{balents1} L. Balents ${\it et\, al.}$, Phys. Rev. B {\bf 71},
144508 (2005); {\it ibid} {\bf 71}, 144509 (2005); L. Balents ${\it
et\, al.}$, Prog. Theor. Phys. Supp., {bf 160}, 314 (2005).

\bibitem{burkov1} A. Burkov and L. Balents, Phys. Rev. B {\bf 72}, 134502
(2005).

\bibitem{sengupta1}K. Sengupta, S. Isakov, and Y.B. Kim, Phys. Rev. B {\bf 72},
134502 (2005).

\bibitem{burkov2}A. Burkov and E. Demler, Phys. Rev. Lett. {\bf 96},
180406 (2006).

\bibitem{squaremonte}G.G. Batrouni {\it et\, al.}, Phys. Rev. Lett. {\bf 74},
2527 (1995); R.T. Scalettar {\it et\, al.}, Phys. Rev. B {\bf 51},
8467 (1995); G.G. Batrouni and R.T. Scalettar, Phys. Rev. Lett. {\bf
84}, 1599 (2000); F. Herbert {it et\, al.}, Phys. Rev. B {\bf 65},
014513 (2002); G. Schmid {\it et\, al.}, Phys. Rev. Lett. {\bf 88},
167208 (2002); A. Kuklov, N. Prokof'ev, and B. Svistunov, Phys. Rev.
Lett. 93, 230402 (2004).

\bibitem{melko1} R. G. Melko {\it et\, al.}, Phys. Rev. Lett. {\bf 95}, 127207
(2005).

\bibitem{damle1} D. Heidarian and K. Damle, Phys. Rev. Lett. {\bf
95}, 127206 (2005).

\bibitem{isakov1} S. V. Isakov {\it et\,al.}, Phys. Rev. Lett. {\bf
97}, 147202 (2006).

\bibitem{damle2}K. Damle and T. Senthil, Phys. Rev. Lett. {\bf 97},
067202 (2006).

\bibitem{wessel1} S. Wessel, cond-mat/0701337 (unpublished).

\bibitem{fisher1} M. P. A. Fisher and D. H. Lee, Phys. Rev. B {\bf 39},
2756 (1989).

\bibitem{tesanovic1} Z. Tesanovic, Phys. Rev. Lett. {\bf 93}, 217004 (2004);
A. Melikyan and Z. Tesanovic, Phys. Rev. B {\bf 71}, 214511 (2005).

\bibitem{dasgupta1} C. Dasgupta and B.I. Halperin, Phys. Rev. Lett. {\bf 47},
1556 (1981)

\bibitem{steve1} M. Wallin, E. Sorensen, A.P. Young and S.M. Girvin,
Phys. Rev. B {\bf 49}, 12115 (1994).

\bibitem{senthil2} C. Lannert, M.P.A Fisher, and T. Senthil,
Phys. Rev. B {\bf 63}, 134510 (2001).

\bibitem{vidal1} J. Vidal ${\it et\,al.}$, Phys. Rev. B {\bf 64},
155306 (2001);  M. Rizzi, V. Cataudella, R. Fazio, Phys. Rev. B {\bf
73}, 144511 (2006).


\bibitem{ye1} L. Jiang and J. Ye, cond-mat/0601083 (unpublished).

\bibitem {sondhi1}  R. Moessner, S.L. Sondhi, and P. Chandra,
Phys. Rev. Lett. {\bf 84}, 4457 (2000).

\bibitem{cabra1} D.C. Cabra {\it et\,al.}, Phys. Rev. B {\bf 71}, 144420
(2005).

\bibitem {damle3} A. Sen, K. Damle, and T. Senthil, cond-mat/0701476.

\bibitem {kuklov1} A. Kuklov {\it et\,al.}, cond-mat/0602466 (unpublished);
A. Kuklov, N. Prokof'ev, and B. Svistunov, cond-mat/0501052
(unpublished).









\end{thebibliography}
\end{document}